\documentstyle[floats,prd,aps,epsfig,eqsecnum,12pt]{revtex}
\makeatletter
\newbox\tempboxa
\newdimen\captionboxsubcount 
\def\capsize#1{\captionboxsubcount=#1pt}
\newdimen\captionboxsub
\captionboxsub=\hsize \advance\captionboxsub by -\captionboxsubcount
\advance\captionboxsub by -\captionboxsubcount
\long\def\@makecaption#1#2{
 \setbox\@tempboxa\hbox{#1 #2}
 \ifdim \wd\@tempboxa >\captionboxsub 
\rightskip=\captionboxsubcount \leftskip=\captionboxsubcount #1 #2 
\else \hbox to\hsize{\hfil\box\@tempboxa\hfil} 
 \fi}
\makeatother
\capsize{30}

\begin{document}
\bibliographystyle{unsrt}
\begin{titlepage}
\begin{flushright}
\begin{minipage}{5cm}
\begin{flushleft}
\small
\baselineskip = 13pt
YCTP-P9-98\\
SU-4240-678\\
hep-ph/9804273 \\
\end{flushleft}
\end{minipage}
\end{flushright}
\begin{center}
\Large\bf
Evidence for a scalar 
$\mbox{\boldmath ${\kappa(900)}$}$ resonance in 
$\mbox{\boldmath $\pi K$}$ scattering 
\end{center}
\vfil
\footnotesep = 12pt
\begin{center}
\large
Deirdre {\sc Black}$^{\it \bf a}$ 
\footnote{Electronic address: {\tt black@physics.syr.edu}}
\quad\quad Amir H. {\sc Fariborz}$^{\it \bf a}$\footnote{Electronic  
address: {\tt amir@suhep.phy.syr.edu}}\\
\vskip 0.5cm
Francesco {\sc Sannino}$^{\it \bf b}$\footnote{
Electronic address : {\tt sannino@apocalypse.physics.yale.edu}} 
\quad\quad
Joseph {\sc Schechter}$^{\it \bf a}$\footnote{
Electronic address : {\tt schechte@suhep.phy.syr.edu}}\\
{\it 
\qquad $^{\it \bf a}$ Department of Physics, Syracuse University, 
Syracuse, NY 13244-1130, USA.} \\
\vskip 0.5cm
{ \it  \qquad $^{\it \bf b}$ 
Department of Physics, Yale University, New Haven, CT 06520-8120,  
USA.}\\
\end{center}
\vfill
\begin{center}
\bf
Abstract
\end{center}
\begin{abstract}
Motivated by the $\displaystyle{{1}/{N_c}}$ expansion, we study a
simple model in which the $\pi K$ scattering amplitude is the sum of a
$\it {current-algebra}$ contact term and resonance pole exchanges.
This phenomenological model is crossing symmetric and, when a putative
light strange scalar meson $\kappa$ is included, satisfies the
unitarity bounds to well above $1$ GeV.  The model also features
chiral dynamics, vector meson dominance and appropriate interference
between the established $K^*_{0}(1430)$ resonance and its predicted
background.  We briefly discuss the physical significance of the
results and directions for further work.
\baselineskip = 17pt
\end{abstract}
\begin{flushleft}
\footnotesize
PACS number(s): 13.75.Lb, 11.15.Pg, 11.80.Et, 12.39.Fe 
\end{flushleft}
\vfill
\end{titlepage}
\setcounter{footnote}{0}
\section{Introduction}
In the present paper we will generalize to the case of $\pi K$
scattering the recent treatment of $\pi\pi$ scattering given in
\cite{Sannino-Schechter,Harada-Sannino-Schechter,Harada-Sannino-Schechter-com}.
There evidence was found to support the existence of a low-mass
relatively broad scalar resonance, denoted $\sigma [m_\sigma =
560$MeV, $\displaystyle{\Gamma_\sigma = 370}$MeV, with pole position
$\displaystyle{s = (0.585 - 0.178i)}$ GeV], in addition to the
well-established scalar $f_0(980)$ resonance. A number of other
authors have also found similar or related results in different models
\cite{Tornqvist,Ishidapi,Morgan-Pennington:93,Janssen-Pearce-Holinde-Speth,lnc,Achasov,Kaminski,Svec,vanBeveren,Scadron,OOP}.

If one accepts a low-lying $\sigma$ and notes the existence of the
isovector scalar $a_0(980)$, as well as the $f_0(980)$, there would be
three scalar resonances below 1GeV.  A great deal of discussion and
controversy over the years has surrounded the issue of the nature of
such very low-mass scalars. The reason is that one expects the
lowest-lying scalars in the quark model to be p-wave $q{\bar q}$ bound
states and hence to have masses comparable to those of the axial and
tensor mesons, already in the 1.2 - 1.6 GeV region (see for example
\cite{Close}).  As an example (see the discussion on page 355 of
\cite{PDG} under the ``Note on Scalar Mesons'') one might form a
conventional scalar nonet from the $f_0(1370)$, $a_0(1450)$,
$K_0{^*}(1430)$ and $f_J(1710)$.  If an assignment like this is
correct it raises the question of why the three scalar candidates
$\sigma$, $f_0(980)$ and $a_0(980)$ are so light, and whether a
general organizing principle for their dynamics can be found.  From
this point of view it is extremely interesting to see if a light
strange scalar resonance, to be denoted $\kappa$, emerges in the study
of $\pi K$ scattering.  Evidence for such a resonance has been found
by some authors - \cite{vanBeveren} using a unitarized
non-relativistic meson model and \cite{Ishida_kappa} using a method of
interfering Breit-Wigner amplitudes with a repulsive background - and
disputed by others - \cite{Tornqvistkappa} using a unitarized quark
model.  The existence of the $\kappa$ would strengthen the point of
view (see for example \cite{Jaffe}) that there is a non-conventional
scalar nonet lying below 1 GeV.

Of course another motivation for studying $\pi K$ scattering using
the approach of \cite {Sannino-Schechter,Harada-Sannino-Schechter} is
to test that approach itself in a context other than $\pi\pi$
scattering.  According to experimental indications \cite{Aston} the
$\pi K$ channel may be a particularly clean one for this purpose in
that the effects of inelastic channels seem to be less important at
moderate energies than for $\pi\pi$ scattering.  Theoretically too,
the $\pi K$ scattering seems cleaner in that its non-trivial quantum
numbers reduce the number of nearby states which can mix with each
other.  This contrasts with the $\pi\pi$ isosinglet channel in which
$(u\bar u + d\bar d)$, $s\bar s$ and glueball states can {\it a priori}
mix.

Perhaps it is useful to remark on the need to ``discover'' a light
scalar meson by an analysis of the sort being undertaken here; why
can't one just rely on an inspection of the phase shifts obtained
directly from experiment?  In the case of the $\pi\pi$ isosinglet
channel, the model of
\cite{Sannino-Schechter,Harada-Sannino-Schechter} for example shows
that the light $\sigma$ is on the broad side and does not dominate its
own channel.  Rather it is only one of three comparable and competing
contributions.  A similar situation is expected and will be seen to
occur for the putative $\kappa$ meson.  Clearly, the reliability of
such a prediction depends on how accurately the ``background'' of the
$\kappa$ can be modeled.  In the present approach that job will be
facilitated by using an effective chiral Lagrangian approach in which
crossing symmetry is manifest.  This insures that important
cross-channel contributions from resonances known to exist in a given
energy region are included.  Furthermore, by using the physical fields
directly, we will not be limiting ourselves to any assumption about a
particular kind of quark substructure for these fields.  This is, on
the one hand, an advantage, since it increases the generality of our
analysis.  On the other hand our demonstration of the need for a
$\kappa$ meson will not immediately answer the interesting question of
what the quark substructure of light scalars is.  In fact, we will not
take a stand on this matter in the present paper and reserve our
speculative notions for elsewhere \cite{black}.

This paper is organized as follows.  In section II there is a brief
review of our approach as it was applied to the $\pi\pi$ scattering problem.
This is used to motivate the specific approximations which we will
make in the present case of $\pi K$ scattering.  Section III treats
the very interesting $J=0$, $\displaystyle {I= \frac {1}{2}}$ channel.
It is shown that postulating the existence of a light $\kappa$-type
resonance enables us to satisfy the unitarity bound in this channel.
In section IV it is further shown that the existence of the $\kappa$
also plays an important role in producing a background phase at the
position of the $K_0^{*}(1430)$ resonance pole; this gives a shape for
the $J=0$, $\displaystyle {I= \frac {1}{2}}$ partial wave amplitude in
agreement with experiment.  The $J=0$, $\displaystyle {I = \frac
{3}{2}}$ channel, which apparently does not contain any exotic
$\displaystyle {I=\frac {3}{2}}$ resonance poles, is studied in section
V. A brief summary and discussion are given in section VI.  For the
reader's convenience, many technical details are compactly assembled
in three Appendices.  Appendix A, B and C are respectively devoted to
scattering kinematics, the underlying chiral Lagrangian and the
``unregularized'' invariant amplitudes.

\section{Review of the Model}
{}For the reader's convenience we will briefly review here the main
features of \cite{Sannino-Schechter,Harada-Sannino-Schechter} in which
$\pi\pi$ scattering was discussed and indicate how they are expected
to generalize to the $\pi K$ case.  For a fuller presentation of the
ideas used, we refer the reader to
\cite{Sannino-Schechter,Harada-Sannino-Schechter}.

The approach is inspired by the $\displaystyle{\frac {1}{N_c}}$
expansion \cite{1n} of QCD.  It is desired to approximate the low
energy (up to the roughly 1 GeV region) part of the leading, order of
$\displaystyle{\frac {1}{N_c}}$, contribution to the meson-meson
scattering amplitude.  It seems to be an outstanding unsolved problem
to obtain an analytic representation of even this leading
contribution.  However, certain of its features \cite{1n} are known.
The amplitude should consist of tree diagrams - contact terms and
resonance exchanges.  Away from the poles (which contain divergences
of the theory in leading order since the resonance widths go as
$\displaystyle{\frac {1}{N_c}}$) the leading order amplitudes are purely
real.  Hence we restrict ourselves to comparing the real parts of our
computed amplitudes with the real parts of the amplitudes deduced from
experiment.

A crucial aspect is the regularization procedure at the s-channel
poles.  The guiding principle is to make the amplitude unitary in the
neighborhood of the pole and the resulting regularization method used
depends on the type of resonance under consideration.  As illustrated
in section II of \cite{Harada-Sannino-Schechter} this gives the
Breit-Wigner prescription for a narrow isolated resonance, a
Breit-Wigner prescription modified by a computed phase shift for a
narrow resonance in a smoothly varying background and a slightly more
general parameterization for the relatively broad light scalar
resonance.

The crossing symmetric amplitude will, to insure chiral symmetry which
works very well near threshold, be computed from the chiral Lagrangian
given in Appendix B (the same one used in
\cite{Sannino-Schechter,Harada-Sannino-Schechter}).  The partial wave
projections of interest will then be obtained according to
(\ref{projected}).  

To see what happens in the case of the $\pi\pi$, $I=J=0$ partial wave
amplitude let us start from threshold and go up in energy.  The
threshold region is well explained by the so-called current
algebra contact term.  However as shown in Fig.~1 of
\cite{Harada-Sannino-Schechter}, this contact amplitude rises rapidly,
already violating the unitarity bound at around $500$ MeV.  It is
postulated that unitarity should be restored by nearby resonance
contributions and this is called ``local cancellation''.  It is also
seen in this figure that the introduction of the $\rho$-meson
contribution markedly improves, but does not completely cure, the
unitarity violation.  However this result makes the possibility of
``local cancellation'' seem plausible.  A certain amount of
experimentation, described in \cite{Sannino-Schechter}, showed that
the remaining violation of the unitarity bound could be neatly cured
by the introduction of a suitably parameterized light scalar $\sigma$
meson.  Figure 9 of \cite{Sannino-Schechter} shows how such a $\sigma$
meson, having a mass close to the energy where the unitarity bound is
violated, kills two birds with one stone.  At lower energies it boosts
the ``current algebra'' result which is slightly too small when
compared with the real part of the experimentally determined
amplitude.  At higher energies it falls rapidly to negative values to
rescue unitarity.  Furthermore in the region of the $f_0(980)$ the
real part of these contributions to the amplitude is brought to zero
which yields a background phase of around $90$ degrees.  In turn (see
section IVA of \cite{Harada-Sannino-Schechter}), this leads to a
Ramsauer-Townsend mechanism \cite{Taylor} which changes the $f_0(980)$
contribution to the cross-section from a peak to the experimentally
observed dip.  All in all a reasonable experimental fit for the
isosinglet scalar amplitude is obtained up to about 1.2 GeV (see
Fig. 4 of \cite{Harada-Sannino-Schechter}).  The great precision of
the chiral perturbation theory \cite{chp} description of the amplitude
very close to threshold has been slightly sacrificed to achieve an
overall description over a considerably larger energy range.

Two additional points can be made.  Investigation of the effect of the
opening of the $\pi\pi \rightarrow K\bar K$ channel (Section V of
\cite{Harada-Sannino-Schechter}) showed that it made a relatively
minor change in the qualitative treatment of $\pi\pi \rightarrow
\pi\pi$ scattering up to about 1.2 GeV.  Amusingly, the same mechanism
for restoring unitarity which worked for $\pi\pi \rightarrow \pi\pi$
seemed also effective for the $\pi\pi \rightarrow K\bar K$, $I=J=0$
amplitude above the $K\bar K$ threshold.  Secondly, it was noted
\cite{Harada-Sannino-Schechter,Sannino-Schechter} that there was a
tendency for contributions from the exchange of the ``next group'' of
resonances - the $f_2(1270)$, the $f_0(1300)$ and the ${\rho}(1450)$ -
to cancel among themselves.  In any event they did not further improve
the fit.  Certainly, in order to carry this treatment still higher in
energy it is necessary to treat the higher resonances more precisely.
In the numerical treatment of
\cite{Harada-Sannino-Schechter,Sannino-Schechter}, it was found that
these effects of inelasticity and the higher resonances could all be
absorbed in relatively minor adjustments of the three parameters used
to describe the light scalar.

\begin{figure}
\centering
\epsfig{file=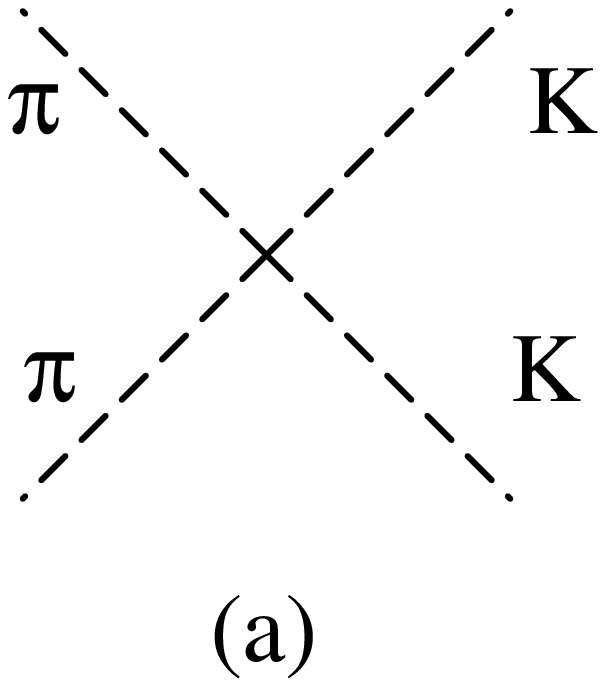, height=1in, angle=0}
\epsfig{file=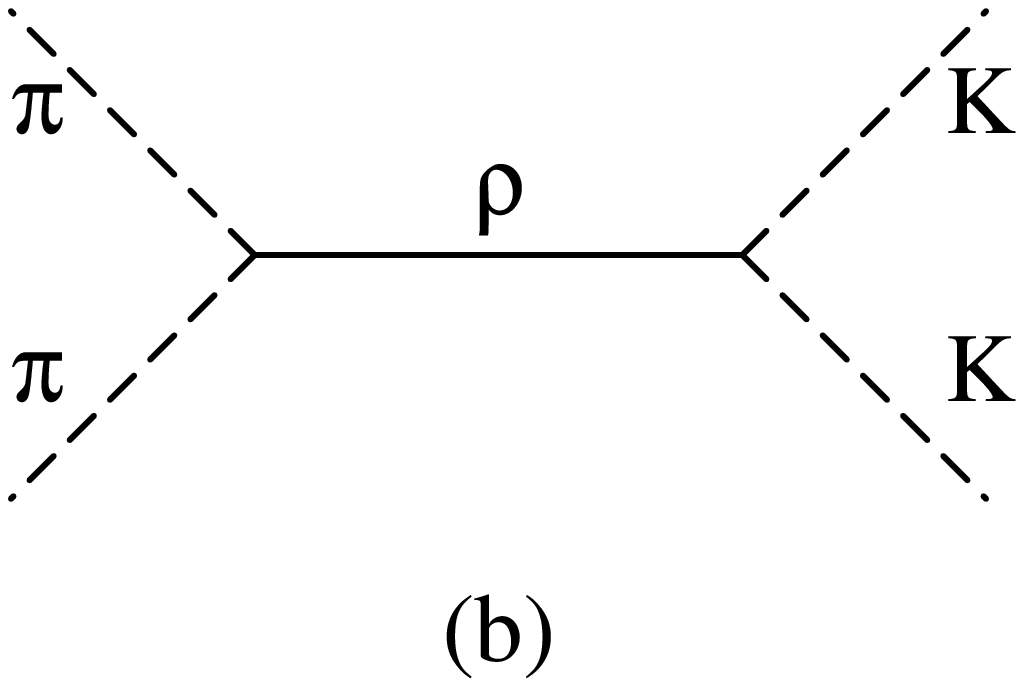, height=1in, angle=0}

\epsfig{file=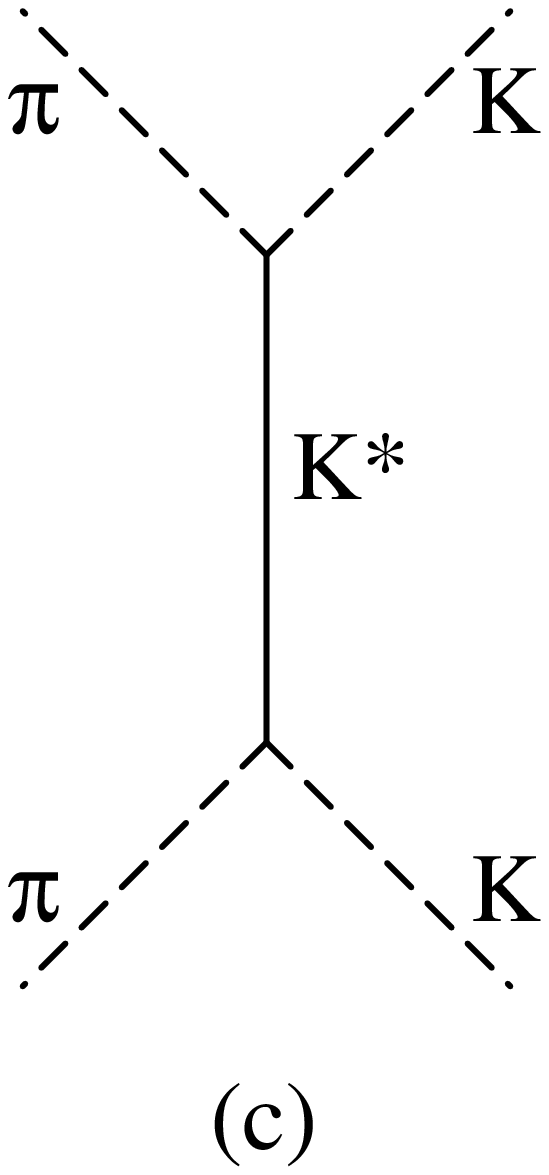, height=2in, angle=0}
\epsfig{file=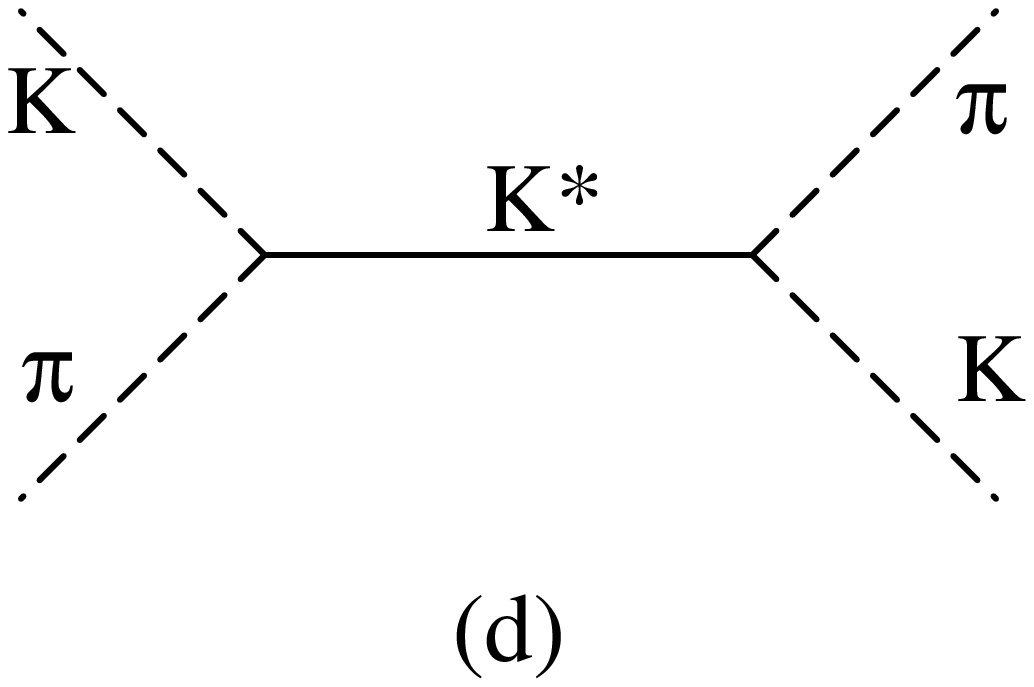, height=1in, angle=0}

\epsfig{file=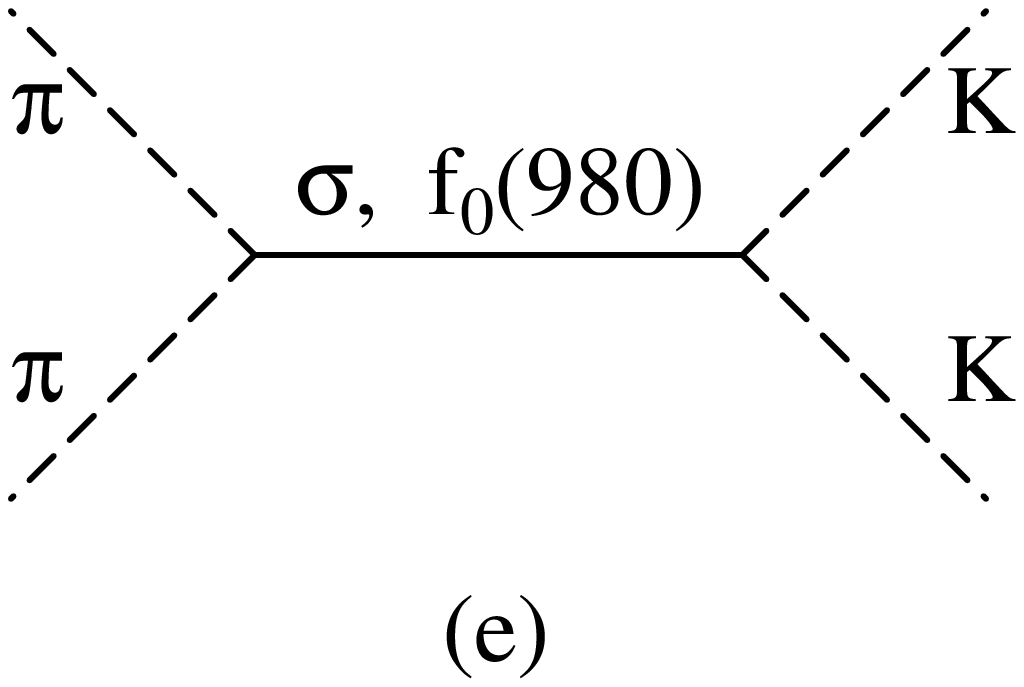, height=1in, angle=0}
\epsfig{file=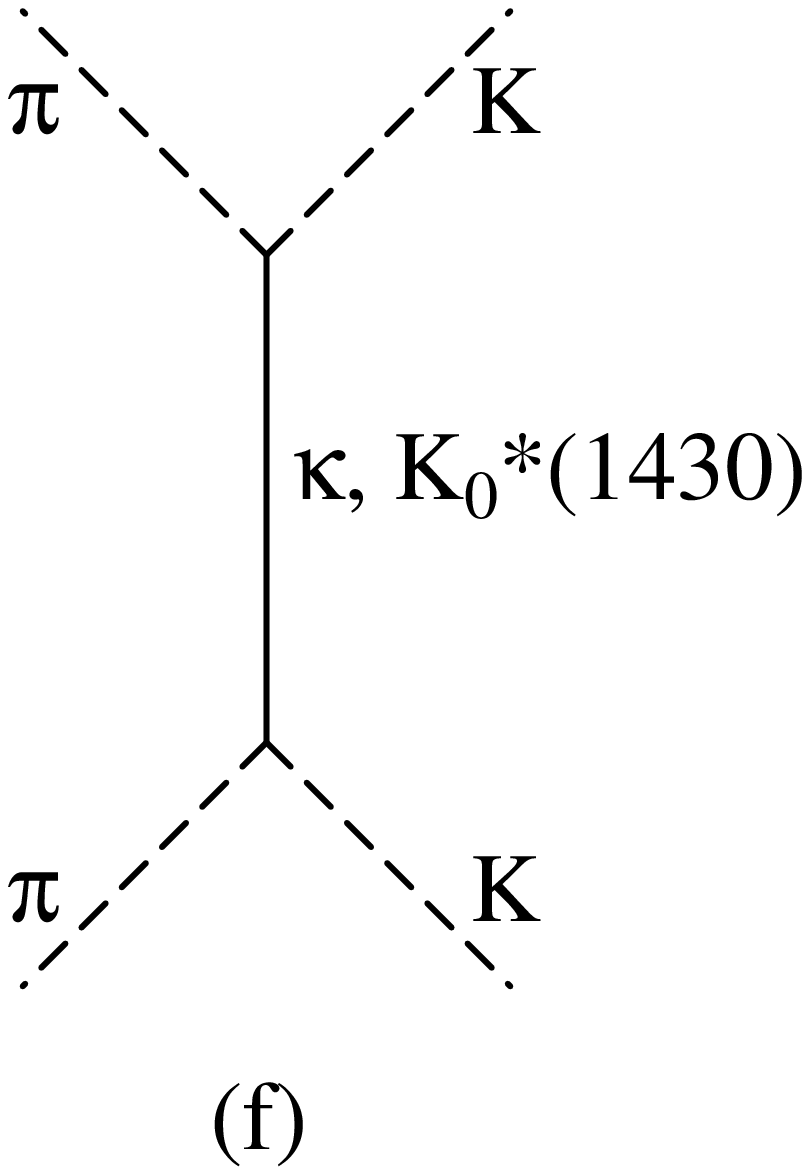, height=2in, angle=0}

\epsfig{file=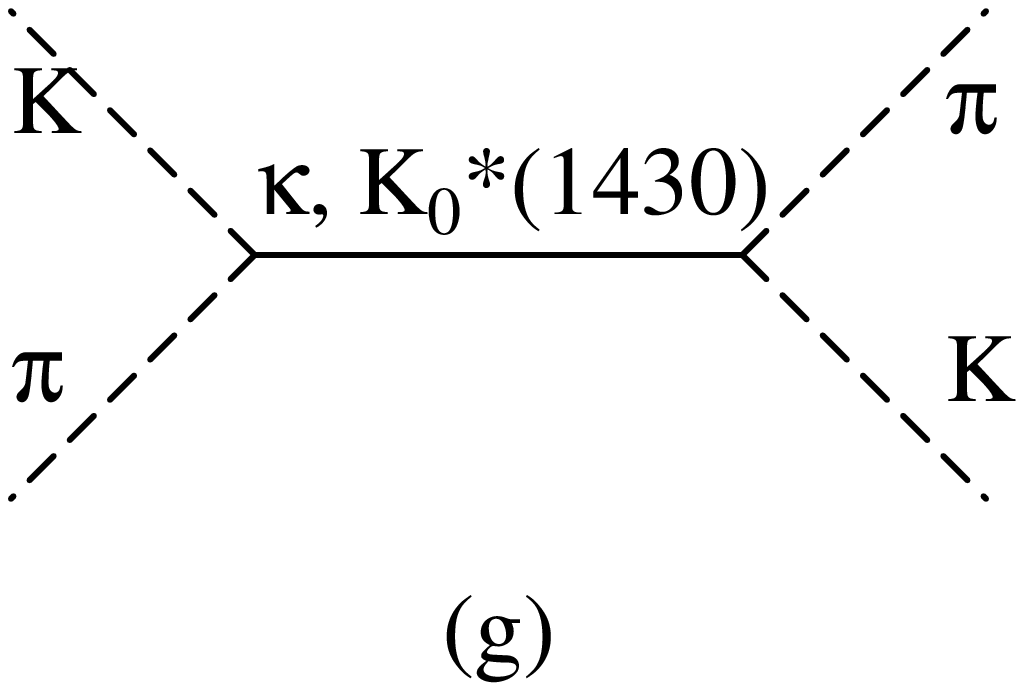, height=1in, angle=0}
\caption
{
Tree diagrams relevant for $\pi K$ scattering in our model
}
\label{FeynD}
\end{figure}

{}From this discussion, it seems that the appropriate model for an
initial study of the generalization to the $\pi K$ case would neglect
the inelastic channels (here $\eta' K$ is apparently \cite{Aston} the
main first one) as well as resonances other than the vector mesons and
the scalars which lie below 1 Gev.  Since we are especially interested
in the $J=0$, $\displaystyle{I={{1}\over 2}}$ channel we will make an
important exception for the $K_0^{*}(1430)$ which has a direct pole in
this channel.  The $K_0^{*}(1430)$ seems to be a reasonable candidate for
an ``ordinary'' p-wave $q\bar q$ scalar.  The diagrams to be
considered are shown in Fig.~\ref{FeynD}.  Notice that a putative light scalar
$\kappa$ has been included.  The main question is whether it is {\it
needed} to satisfy the unitarity bound.  Actually our treatment of the
$\displaystyle{I={{1}\over 2}}$ channel turns out to be conceptually
similar to the experimental analysis of \cite{Aston}.  They
parameterize the $\displaystyle{I={{1}\over 2}}$, $J=0$ channel amplitude by an
effective range background piece plus a modified Breit-Wigner term for
the $K_0^{*}(1430)$.  We work from our crossing symmetric invariant
amplitude, so in effect their background corresponds to the sum of all
our diagrams, except for the $K_0^{*}(1430)$ pole terms in
Fig.~\ref{FeynD}.  Since their parameters for the $K_0^{*}(1430)$ are determined
by this method we choose to fit the $K_0^{*}(1430)$ and $\kappa$
parameters simultaneously.

\section{evidence for the scalar $\kappa$(900) in the
$\displaystyle{I= \frac{1}{2}}$ channel}

\begin{figure} 
\centering
\epsfig{file=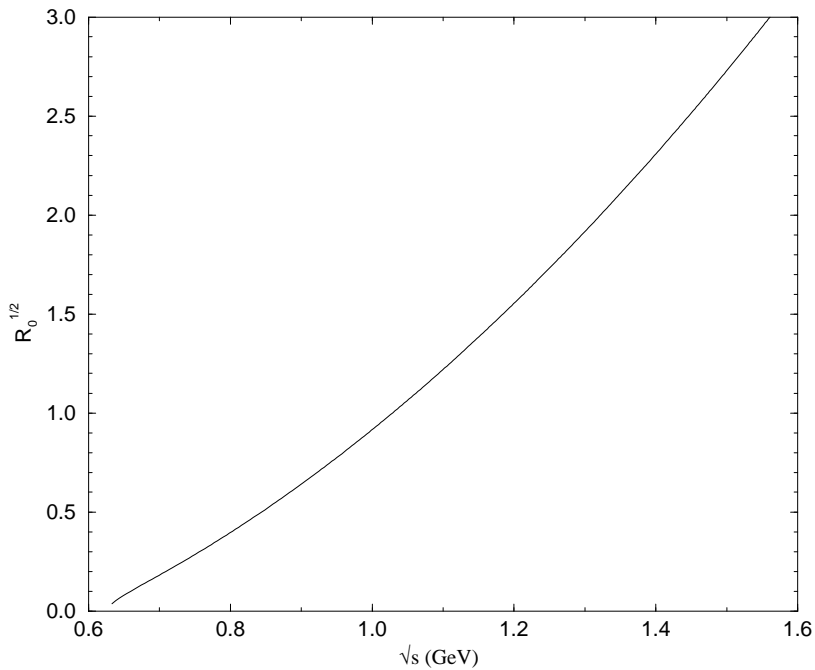, height=5in, angle=270}
\caption
{
Current algebra contribution to
$R^{1/2}_0$.
}
\label{fig-CA}
\end{figure}

In this section we make an initial study of the $\displaystyle{I=\frac
{1}{2}}$ and $J=0$ projection of the real part of the $\pi K$
scattering amplitude $T_0^{1/2}$ defined in (\ref{projected}). As in
the $\pi\pi$ case we start with the well-known ``current algebra''
amplitude.  This can be calculated from the second term of the
Lagrangian (\ref{Lag: sym}) together with (\ref{pi mass 1}).  If the
vector mesons are not included in this chiral Lagrangian, then this is
the same as using the more conventional chiral Lagrangian, including
only pseudoscalars \cite{Cronin}:
\begin{equation}
{\cal L}_1=
- { {F_\pi^2} \over 8} {\rm Tr} 
\left( \partial_\mu U \partial_\mu U^\dagger  \right)
+{\rm Tr}
\left[
       {\cal B} \left( U + U^\dagger \right)
\right],
\label{L1}
\end{equation}
in which $\displaystyle{U=e^{2i{\phi\over F_\pi}}}$, with $\phi$ the
$3 \times 3$ matrix of pseudoscalar fields and $F_\pi = 132 $ MeV the
pion decay constant. ${\cal B}$ is a diagonal matrix $(B_1, B_1, B_3)$
with $B_1=m_\pi^2 F_\pi^2/8=B_2$ and
$B_3=F_\pi^2(m_K^2-m_\pi^2/2)/4$. This is the dominant minimal
symmetry breaking term for the pseudoscalar mesons. We shall choose
$m_\pi=137$ MeV and $m_K=496$ MeV.  Using (\ref{CA-3/2}) together with
(\ref{1/2}), gives the $\displaystyle{I=\frac {1}{2}}$ invariant
amplitude
\begin{equation}
A^{1/2}_{CA}(s,t,u)={1\over {2 F_\pi^2}}\left[ 2(s-u) + t \right],
\label{T12CA_stu}
\end{equation}
and we will refer to this as the {\it current algebra} result.  Using
(\ref{projected}) we find the $J=0$ partial wave amplitude to be:
\begin{equation}
{R_0^{1/2}}_{CA}=
{
   {q} 
    \over 
   {8\pi F_\pi^2 \sqrt{s} }
}
\left[
       2 (s-m_\pi^2-m_K^2)- 3 q^2
\right],
\end{equation}
where the magnitude of the center of mass momentum q(s) is given in
(\ref{kinematical}).  The current algebra result is shown in
Fig.~\ref{fig-CA}, indicating a severe violation of the unitarity
bound (\ref{unitarity}) beyond approximately $900$ MeV.  This
resembles the violation of the unitarity bound by the current algebra
prediction in the $\pi\pi$ case.  As in that case we will try to solve
this problem by including resonance contributions to the scattering
amplitude.
  
First consider the effect of the vector mesons.  There are $\rho$ and
$K^*$ exchanges and a direct $K^*$ pole as illustrated in Figs~1(b), 1(c) and
1(d).  The relevant coupling constants are read off from the
$\rho_{\mu}v_{\mu}$ piece in the first term of (\ref{Lag: sym}).
Symmetry breaking contributions are small
\cite{Schechter-Subbaraman-Weigel} and will be neglected here.  As an
example, the invariant amplitude representing the two $K^*$
diagrams is
\begin{equation}
A^{1/2}_{K^*} = {\frac {3}{2}} {P\left(u,t,s\right)} - {\frac
{1}{2}} {P\left(s,t,u\right)},
\end{equation}
with
\begin{equation}
P\left(u,t,s \right) =  \frac {g_{\rho\pi\pi}^2}{4m_{K^*}^2} \left[\frac
{m_{K^*}^2\left(t-u\right) + {\left(m_K^2 - m_\pi^2\right)}^2}{m_{K^*}^2 - s
- im_{K^*}\Gamma_{K^*}\theta \left(s-s_{th}\right)}\right],
\label{regularised}
\end{equation}
where $\Gamma _{K^*}$ is the $K^*$ width, $s_{th}={\left( m_k + m_\pi
\right)}^2$, $\theta$ is the Heaviside step function and we take
$m_\rho=0.77 $ GeV, $g_{\rho\pi\pi}=8.56$ and $m_{K^*}=0.89$ GeV.  We
have added a conventional width term in order to regularize the
s-channel pole.  We may more generally regard this regularization as
the imposition of unitarity on the $J=0$, $\displaystyle{I=\frac
{1}{2}}$ partial wave amplitude in the region near the $K^*$ mass.
Comparison with (\ref{1/2}) shows that this regularization formally
maintains crossing symmetry.  Actually our results are not very
sensitive to the fine details of the regularization function.

\begin{figure} 
\centering
\epsfig{file=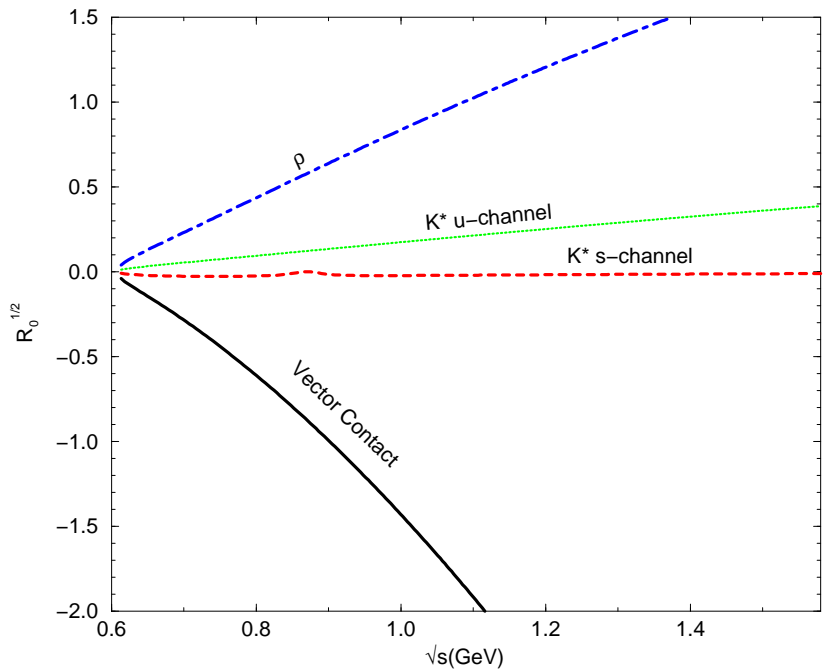, height=5in, angle=270}
\caption
{ Individual vector contributions to $R^{1/2}_0$.  }
\label{fig-vectors}
\end{figure}

\begin{figure} 
\centering
\epsfig{file=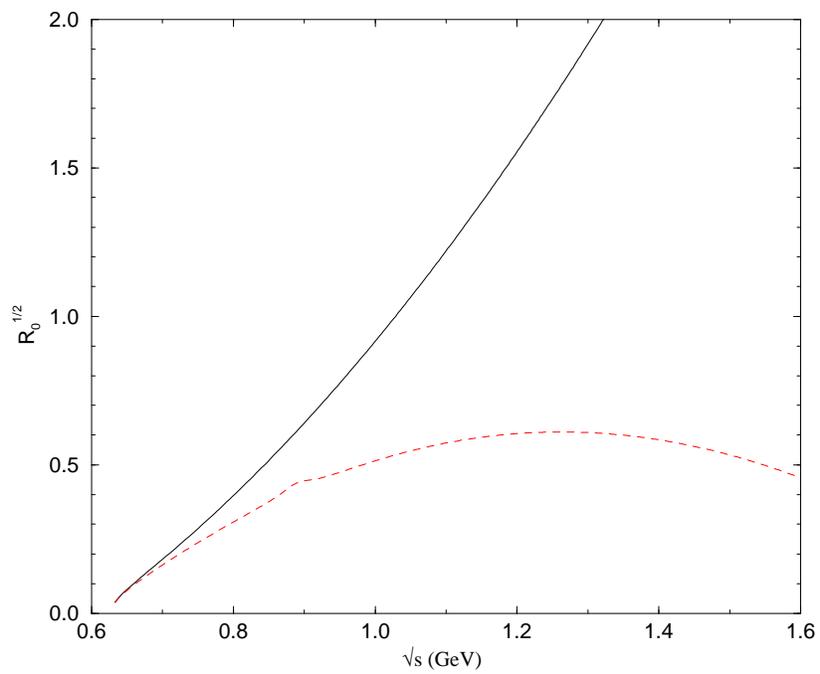, height=5in, angle=270}
\caption
{
Contribution of current algebra (solid line), and  
current algebra + vectors (dashed-line) to 
$R^{1/2}_0$.
}
\label{fig-CAV}
\end{figure}

The contributions associated with the vector mesons including the
$\rho$ exchange diagram, the $K^*$ diagrams and a new contact term
arising from the $v_{\mu}v_{\mu}$ piece in (\ref{Lag: sym}) are
plotted \footnote{The bump in the s-channel $K^*$ contribution arises
because the amplitude is forced to $\it {rise}$ to zero at the $K^*$
mass by the spin 1 projection property of the $K^*$ propagator.}in
Fig.~\ref{fig-vectors}.  As expected, the direct contribution due to
the s-channel $K^*$ pole upon projection into the scalar channel is
almost zero.  In fact it is the new contact term which is seen to play
a crucial role in helping to restore unitarity.  This term is negative
and thus balances the positive current algebra piece.  It arises as a
consequence of casting the Lagrangian with vectors (\ref{Lag: sym}) in
a chirally invariant form.  The effect of all the vector
contributions, added to the current algebra piece is displayed in
Fig.~\ref{fig-CAV}.  It can be seen that, while individual terms
violate the unitarity bound, the introduction of vectors has pulled
the curve down so that it almost lies within the bound.  A similar
improvement, due to the inclusion of vectors, was observed in the
analysis of the $\pi\pi$ scattering amplitude
\cite{Sannino-Schechter,Harada-Sannino-Schechter}.
  
So far we have not used any unknown parameters, so the current algebra
and vector contributions are fixed.  Actually the violation of
unitarity is smaller than at the corresponding stage of the analogous
$\pi\pi$ calculation and one might be inclined to stop at this point.
However, in our framework, we should include other diagrams for
resonances lying within the energy range of interest.  There is the
established $f_0(980)$ as well as the $\sigma (560)$ which should be
included for self-consistency.  Of course the role played by a
possible strange scalar is of great interest.  The relevant Feynman
diagrams are shown in Figs~1(e), 1(f) and 1(g).  Another reason for
inclusion of these resonances can be seen by looking ahead to the
experimentally deduced form for $R_0^{1/2}$
(Fig.~{\ref{fig-expdata}}).  The sharp dip near 1400 MeV could not be
explained from the total current algebra plus vector amplitude of
Fig.~{\ref{fig-CAV}}.

In order to compute the scalar exchange diagrams we need the following
pieces of the scalar-pseudoscalar-pseudoscalar interaction
Lagrangian given at the end of Appendix B:

\begin{eqnarray}
{\cal L}_{scalars}=&-&{\sqrt{2}}{\gamma_{\sigma \pi \pi}}\left( 
\sigma \partial_{\mu} {\pi ^+} \partial_{\mu} {\pi^-} + .... \right) -
\frac{\gamma_{\sigma K \bar K}}{\sqrt{2}} \left( 
\sigma \partial_{\mu}K^{+} \partial_{\mu}K^{-} + .... \right) \nonumber \\
&-&{\sqrt{2}}{\gamma_{f_0\pi \pi}} \left( f_0 \partial_{\mu} {\pi ^+}
 \partial_{\mu} {\pi ^-} + .... \right) \
- \frac{\gamma_{f_0 K\bar K}}{\sqrt{2}} \left( 
f_0 \partial_{\mu}K^{+} \partial_{\mu}K^{-} + .... \right) \nonumber\\
&-& {\gamma_{\kappa K\pi}} \left( 
\kappa^0\partial_{\mu} K^{-} \partial_{\mu} \pi^+ + {.....} \right) \ .
\label{Scalar Lagrangian}
\end{eqnarray}
For generality we are not assuming any model to relate these couplings
to each other.  Furthermore, as discussed in Appendix B, the
derivative coupling is the one which would follow from a chiral
invariant model.  Also, the terms shown are the particular ones needed
to compute the required $\pi ^+ K^+$ scattering amplitude in
(\ref{3/2}).  The coupling constants $\gamma_{\sigma\pi\pi}$,
$\gamma_{f_0 \pi\pi}$ and $\gamma_{f_0 K \bar K}$ were estimated in
\cite{Harada-Sannino-Schechter}:

\begin{equation}
\left| \gamma_{\sigma\pi\pi} \right| = 7.81 \; {\rm GeV}^{-1}, \quad \quad
\left|\gamma_{f_0 \pi\pi} \right| = 2.43 \; {\rm GeV}^{-1}, \quad \quad \left|
\gamma_{f_0 K \bar K} \right| = 10 \; {\rm GeV}^{-1}.
\label{couplings}
\end{equation}
Of the needed $\sigma$ and $f_0(980)$ coupling constants, only
$\gamma_{\sigma K \bar K}$ was deduced using $SU(3)$
invariance in some way (which implies specializing to a given quark
substructure for the scalars).  In our final analysis we will
thus, for generality, consider the effect of varying the magnitude and sign of
$\gamma_{\sigma K \bar K}$.  Because the $f_0(980)$ contribution is
rather small, the relative sign of $\gamma_{f_0
\pi\pi}$ and $\gamma_{f_0 K \bar K}$ is of less interest.  

\begin{figure} 
\centering
\epsfig{file=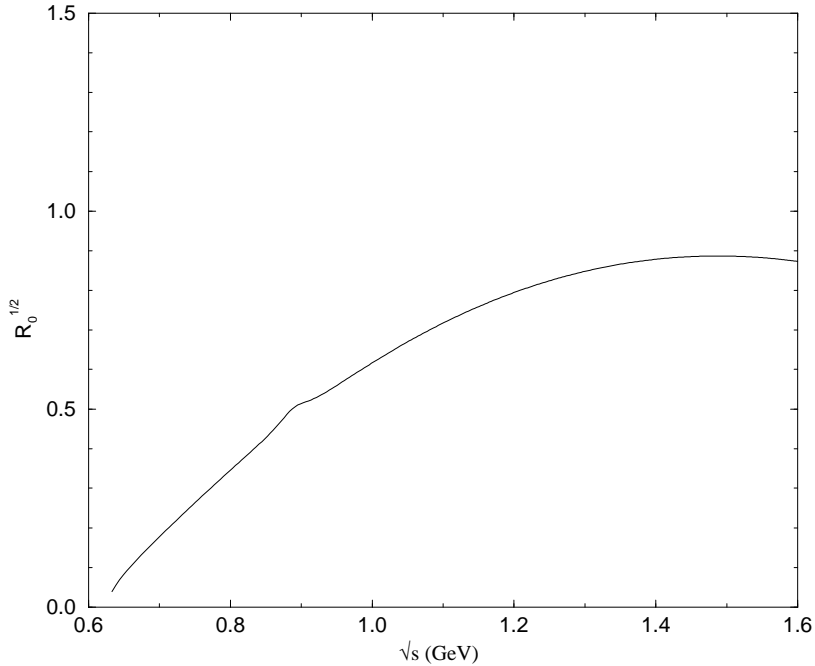, height=5in, angle=270}
\caption
{
Current algebra + vectors + $\sigma$ + $f_0(980)$ contribution to
$R^{1/2}_0$.
}
\label{fig-CAVSF}
\end{figure}

Firstly, we take into account the $\sigma$-meson and the
well-established $f_0(980)$.  Using (\ref{sigma-3/2}) and (\ref{1/2})
we find the $\sigma$ contribution to the invariant amplitude to be 
\begin{equation}
A^{1/2}_\sigma(s,t,u)= { {\gamma_{\sigma\pi\pi}\gamma_{\sigma K{\bar
K}} }\over 4} { {(t-2m_\pi^2)(t-2m_K^2)} \over {m_\sigma^2-t} }.
\label{sigma-1/2}
\end{equation}  
The $f_0(980)$ amplitude has an identical structure with $\sigma
\rightarrow f_0$ everywhere.  We shall take $m_\sigma=0.55$ GeV and
$m_{f_0} = 0.98$ GeV.  For now we take $\gamma_{\sigma K \bar K} =
\gamma_{\sigma\pi\pi}$ and $\gamma_{f_0 K \bar K} \gamma_{f_0 \pi\pi}$
to be positive.  Then a plot showing the effect of adding the
projection of (\ref{sigma-1/2}) into the scalar partial wave channel
is given in Fig.~{\ref{fig-CAVSF}}.  Both the $\sigma$ and $f_0(980)$
contributions are positive, but that of the $\sigma$ is roughly three
times larger.  It is clear that these contributions make the unitarity
violation slightly worse.
 
\begin{figure} 
\centering
\epsfig{file=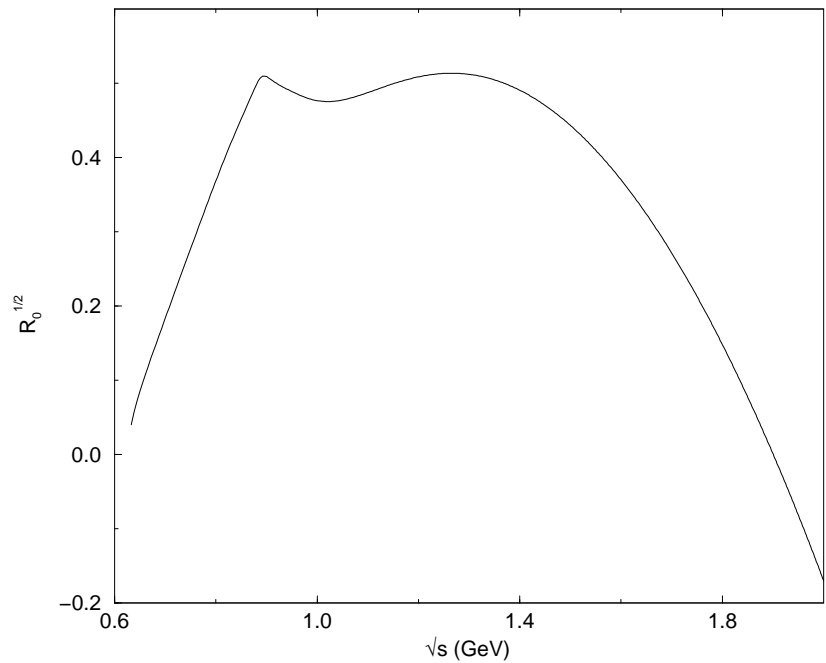, height=5in, angle=270}
\caption
{
Contribution of current algebra + vectors + $\sigma$ +$f_0(980)$ + $\kappa$ to
$R^{1/2}_0$ for $\kappa$ parameters quoted in section III of text.
}
\label{fig-CAVSFK}
\end{figure}

Now let us consider the strange scalar $\kappa$ contribution.  Its
regularized $\displaystyle{I=\frac{1}{2}}$ invariant amplitude is
similarly found to be:
\begin{equation}
A^{1/2}_{\kappa}\left(s,t,u\right)=\frac{\gamma^2_{\kappa K\pi}}{8}\left[ 
\frac{3\left( s - m^2_{\pi} - m^2_{K}\right)^2}{m^2_{\kappa} - s - 
im_{\kappa} G'_{\kappa} \theta \left(s - s_{th} \right) } - \frac{\left( u - m^2_{\pi} -
m^2_{K}\right)^2}{m^2_{\kappa} - u - im_{\kappa} G'_{\kappa} \theta \left( u -
s_{th} \right)} \right].
\label{kappa-1/2}
\end{equation}
As for the $K^*$, this regularization is formally crossing symmetric
(the u-channel regularization term will vanish in the physical
region).  We will treat $m_\kappa$, $\gamma _{\kappa K\pi}$ and
$G'_\kappa$ as independent parameters.  Analogously to the treatment
of the light broad $\sigma (560)$, we have introduced a possible
deviation from the pure Breit-Wigner form by allowing $G'_\kappa$ to
be a free parameter.  The first term in (\ref{kappa-1/2}) is a direct
channel pole and should be extremely important at energies around
$m_\kappa$.  Thus, as in the $\pi\pi$ case it may be used to cure the
unitarity violation of the $J=0$ partial wave amplitude.  Since the
real part of a direct channel resonance contribution turns sharply
negative just above the resonance energy and the graph in
Fig.~{\ref{fig-CAVSF}} rises above the positive unitarity bound at
around $900$ MeV we are led to choose $m_\kappa$ to lie roughly around
this energy.  With the additional illustrative choices $\gamma_{\kappa
K \pi} = 4.8$ GeV$^{-1}$ and $G'_{\kappa} = 280$ MeV we see from
Fig.~{\ref{fig-CAVSFK}}, which is a plot of $R_0^{1/2}$ including also
the contribution of the $J=0$ partial wave projection of
(\ref{kappa-1/2}), that it is easy to achieve a fit in which the
unitarity bound is roughly satisified.  The parameters chosen above
will be seen in the next section to be close to those needed for a fit
to the experimental data.

We obtain the deviation of our $\kappa$ parameterization from a pure
Breit-Wigner shape by noting that near the resonance the $J=0$
partial wave projection of (\ref{kappa-1/2}) is:
\begin{equation}
\frac {m_{\kappa} G_{\kappa}}{m_{\kappa}^2 - s -
im_{\kappa}G'_{\kappa}},
\end{equation}
where the perturbative width $G_{\kappa}$ is given by
\begin{equation}
G_{\kappa}= \frac {3 {\gamma}^2_{\kappa K \pi} q(m^2_{\kappa})}{64 \pi
m^2_{\kappa}} \left( m^2_\kappa - m^2_K - m^2_{\pi}
\right)^2,
\end{equation}
and $q(m^2_\kappa)$ is defined in (\ref{kinematical}).
$\displaystyle{\frac {G_{\kappa}}{G'_\kappa} = 1}$ is the pure
Breit-Wigner situation.  The result $\displaystyle{\frac
{G_{\kappa}}{G'_\kappa} = 0.13}$ is similar to $\displaystyle{\frac
{G_{\sigma}}{G'_\sigma} = 0.29}$ which was previously obtained
\cite{Sannino-Schechter,Harada-Sannino-Schechter} for the $\sigma$.
It seems that such deviations for the low mass scalars are a
characteristic feature of our model.  Ordinarily, when the resonance
is a dominant feature by itself, the Breit-Wigner form may be regarded
as equivalent to unitarity near the resonance.  However, in our model,
there are several different interfering contributions in the low mass
region and all work together to keep the partial wave amplitude within
the unitarity bound.

\section{Global Fit to data in the $\displaystyle{J=0}$, 
$\displaystyle{I=\frac{1}{2}}$ Channel}
The magnitude and phase of the experimental
$\displaystyle{I=\frac{1}{2}}$ s-wave amplitude are given in Fig.~15
of Aston et al \cite{Aston}, based on a high statistics study of the
reaction ${K^-}p \rightarrow {K^-}{\pi ^+}n$.  We have translated these
to the real part ${R^{1/2}_0}\left(s\right)$, which is required for
our approach, and show the results\footnote{Our error bars are based
on propagating the errors in \cite{Aston}, assuming conservatively
these in turn to be given by the experimental circles in
Fig.~15 of \cite{Aston}.} in Fig.~\ref{fig-expdata}.  It is clear that
when one looks at the real part there is an interesting dip at around
$1400$ MeV.  This is explained as the relatively narrow strange scalar
resonance ${K^*_0}(1430)$, which is generally considered to be the
best candidate for a p-wave $q \bar q$ state.  From our point of view
the most interesting question is whether our model including the
$\kappa$ meson provides the correct background structure to explain
the overall shape of $R_0^{1/2}$ in this region.  The role of the
${K^*_0}(1430)$ thus seems analogous to that of the $f_0(980)$ in the
$I=J=0$ partial wave amplitude for $\pi\pi$ scattering.

In that case, as mentioned in section I, the interplay between the
narrow resonance with its background was introduced as a
regularization of the direct channel resonance pole which is
$\displaystyle{\propto {\frac {1}{s - m_*^2}}}$.  In the vicinity of
the resonance, upon projection into the appropriate partial wave, one
sets the amplitude equal to
\begin{equation}
\frac {e^{2i\delta}m_* \Gamma_*}{m_*^2 - s - im_* \Gamma_*} +
e^{i\delta} {\rm sin}{\delta},
\label{total amplitude}
\end{equation}
where $m_*$ and $\Gamma_*$ are the resonance mass and width, while
$\delta$ is the background phase which is assumed to be constant in
the neighborhood of the resonance.  This form automatically makes the
amplitude unitary in this region.  We took our total calculated
amplitude (which was crossing symmetric), without the $f_0(980)$
contribution, evaluated at the position of the resonance, to be the
second term in (\ref{total amplitude}); this allowed us to interpret
the invariant amplitude (\ref{total amplitude}) as being formally
crossing symmetric.  

\begin{figure} 
\centering
\epsfig{file=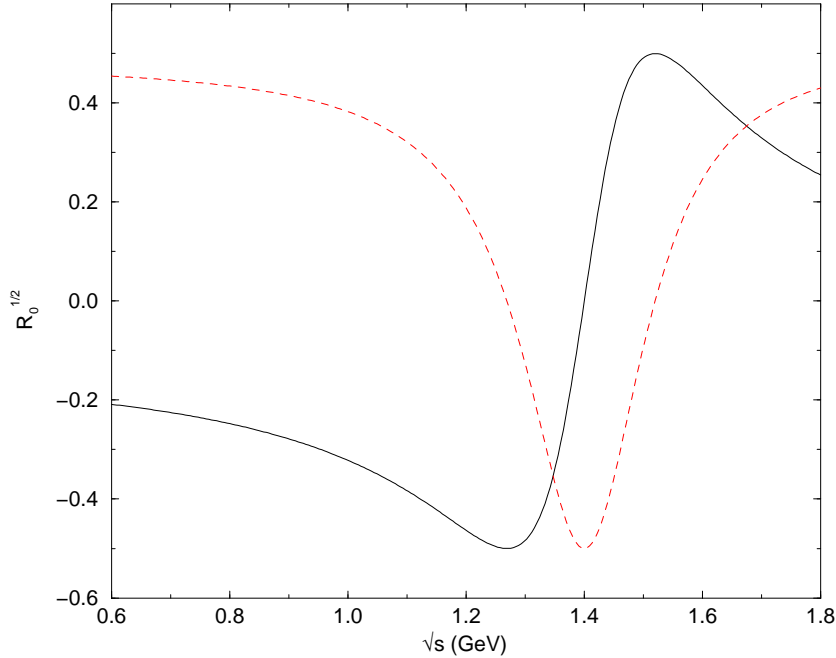, height=5in, angle=270}
\caption
{Shape of $R^{1/2}_0$ derived from Eq. (\ref{total amplitude}) for
resonance ($m=1.4$ GeV and $\Gamma = 0.25$ GeV) in the presence of a
background.  Plot shows two choices for the background phase - $\delta
_{BG} = \frac {\pi}{2}$ (solid line) and $\delta _{BG} = \frac {\pi}{4}$
(dashed line).}
\label{fig-resonance}
\end{figure}

\begin{figure}
\centering
\epsfig{file=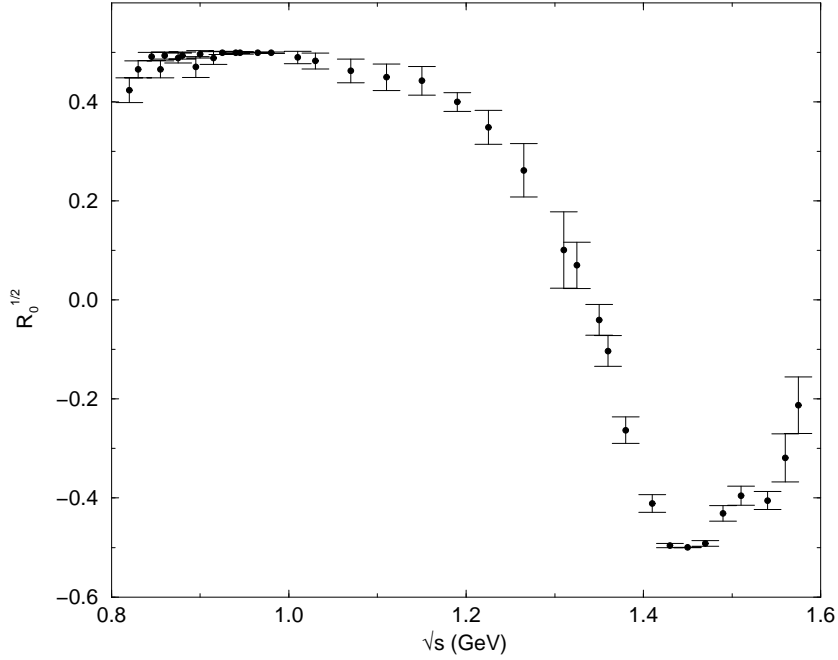, height=5in, angle=270}
\caption
{
Experimental Data for $R^{1/2}_0$.
}
\label{fig-expdata}
\end{figure}

It turns out that there is an interesting
difference between the $\pi\pi$ and $K \pi$ situations.  This can
easily be seen by focusing on the real part of (\ref{total
amplitude}) which is:
\begin{equation}
\frac {1}{2} {{\rm sin} {2\delta}} + \frac{m_* \Gamma _*}{{\left(m_*^2 -
s\right)}^2 + m_*^2 \Gamma _*^2} \left[ \left( m_*^2 - s \right)
{\rm cos}{2\delta} - m_* \Gamma_* {\rm sin} {2\delta} \right].
\label{Re(total amplitude)}
\end{equation}
The shape of this curve depends on the value of $\delta$.  In the
$\pi\pi$ case, the background naturally produced a phase
$\displaystyle{\delta \approx {\pi \over 2}}$ at the position of the
$f_0(980)$.  This yields the shape indicated in 
Fig.~{\ref{fig-resonance}} which just amounts to a sign reversal of the
usual resonance function (in the absence of a background) - the
Ramsauer-Townsend mechanism \cite{Taylor}.  On the other hand,
Fig.~\ref{fig-CAVSFK} shows that $R_0^{1/2}$ is almost
$\displaystyle{\frac{1}{2}}$ at around $1400$ MeV, so that we expect
to have $\displaystyle{\delta \approx {\pi \over 4}}$.  This gives the
other shape shown in Fig.~\ref{fig-resonance} which, in fact, basically
agrees with the experimental $K\pi$ channel picture in
Fig.~{\ref{fig-expdata}}.

Now let us consider the detailed application of this mechanism to
$K\pi$ scattering.  The contribution of the $K_0^*(1430)$ to the
$\displaystyle{I=\frac{1}{2}}$ channel is structurally similar to that of the $\kappa$ 
in (\ref{kappa-1/2}).  The real part of this contribution to the
regularized invariant amplitude is 
\begin{eqnarray}
{\rm Re}\left[ A^{1/2}_{*}\left(s,t,u\right)\right] &=&
\frac{{\gamma_*}^2}{8} {\rm Re} \left[ {e^{2i\delta \theta \left(
s-s_{th}\right) }}\frac{3{\left( s - m^2_{\pi} - m^2_{K}\right)}
^2}{m^2_* - s - im_{*} G'_{*} \theta \left( s - s_{th} \right)
}\right] \nonumber \\ &-& \frac{{\gamma_*}^2}{8} {\rm Re}\left[
{e^{2i\delta\theta\left( u-s_{th}\right) }}\frac {{\left( u -
m^2_{\pi} - m^2_{K}\right)} ^2}{m^2_{*} - u - im_{*} G'_{*} \theta
\left( u - s_{th} \right)} \right].
\label{K(1430)-1/2}
\end{eqnarray}
Here we have denoted quantities associated with the $K_0^*(1430)$ by a
star subscript.  In particular, $m_*$ is now the mass of the
$K_0^*(1430)$.  The quantity $\gamma_*$ is defined in terms of the
$K_0^*(1430)$ partial width into $K\pi$ by:
\begin{equation}
\Gamma \left( K_0^*(1430) \rightarrow K\pi \right) = \frac
{3{\gamma_*}^2 q(m_*^2)(m_*^2 - m_{\pi}^2 - m_K^2)}{64\pi m_*^2},
\label{K(1430) partial width}
\end{equation}
where $q(s)$ is defined in (\ref{kinematical}).  The background
phase $\delta$ will not be considered an arbitrary parameter but shall
be the constant quantity defined from
\begin{equation}
\frac{1}{2} {\rm sin} {2\delta} = \widetilde
R_0^{1/2}(s=m_*^2),
\label{definition of background phase}
\end{equation}
where $\widetilde R_0^{1/2}\left(s\right)$ is the real part of the
partial wave amplitude previously comuted as the sum of the crossing symmetric
current algebra, vector, $\sigma$, $f_0(980)$ and $\kappa$ pieces
found in section III.  With these arrangements the total invariant
amplitude is formally crossing symmetric.  In order to see the
connection with the unitary form (near the resonance) in (\ref{total
amplitude}) and (\ref{Re(total amplitude)}), we simply note that the
second term in (\ref{K(1430)-1/2}) is numerically dominated by the
first term which contains a pole in the physical region. Finally, for
the sake of generality, we shall consider $G'_*$ to be a fitting
parameter, not necessarily equal to $\Gamma \left( K_0^*(1430)
\rightarrow K\pi \right)$.  This allows for the possibility of some
inelasticity.

We notice that the mechanism shown in (\ref{total amplitude})
implicitly demands a background which does not violate the unitarity
bounds at the resonance mass $m_*$.  This provides a justification for
the existence of the $\kappa$ meson, as we showed in the last section
that it is needed to restore unitarity (compare Fig.~{\ref{fig-CAVSF}}
and Fig.~{\ref{fig-CAVSFK}}).

\begin{figure}
\centering
\epsfig{file=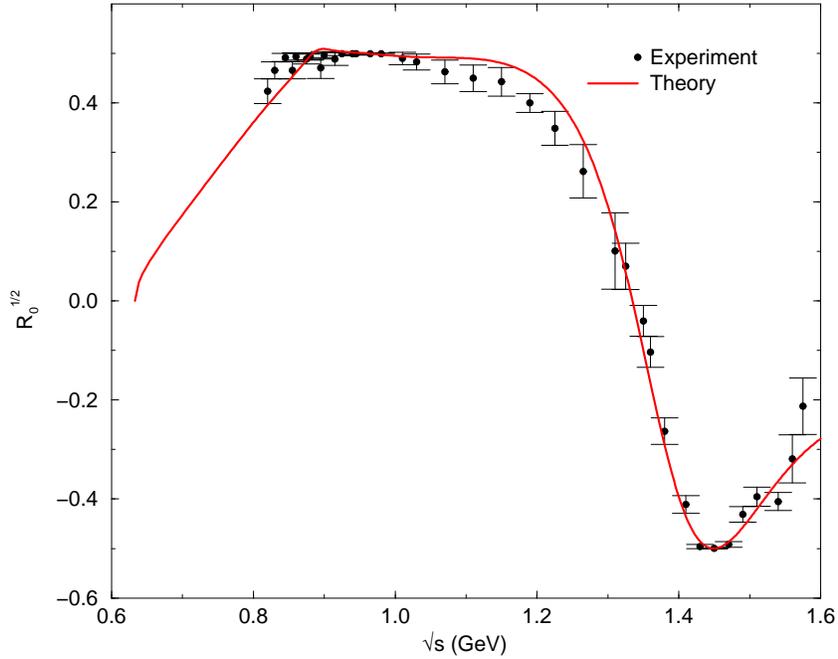, height=5in, angle=270}
\caption
{
Comparison of the theoretical prediction of $R^{1/2}_0$ with its
experimental data (for choice $\gamma_{\sigma K\bar K}=\gamma_{\sigma\pi\pi}$)
}
\label{fig-fit}
\end{figure}

\begin{figure}
\centering
\epsfig{file=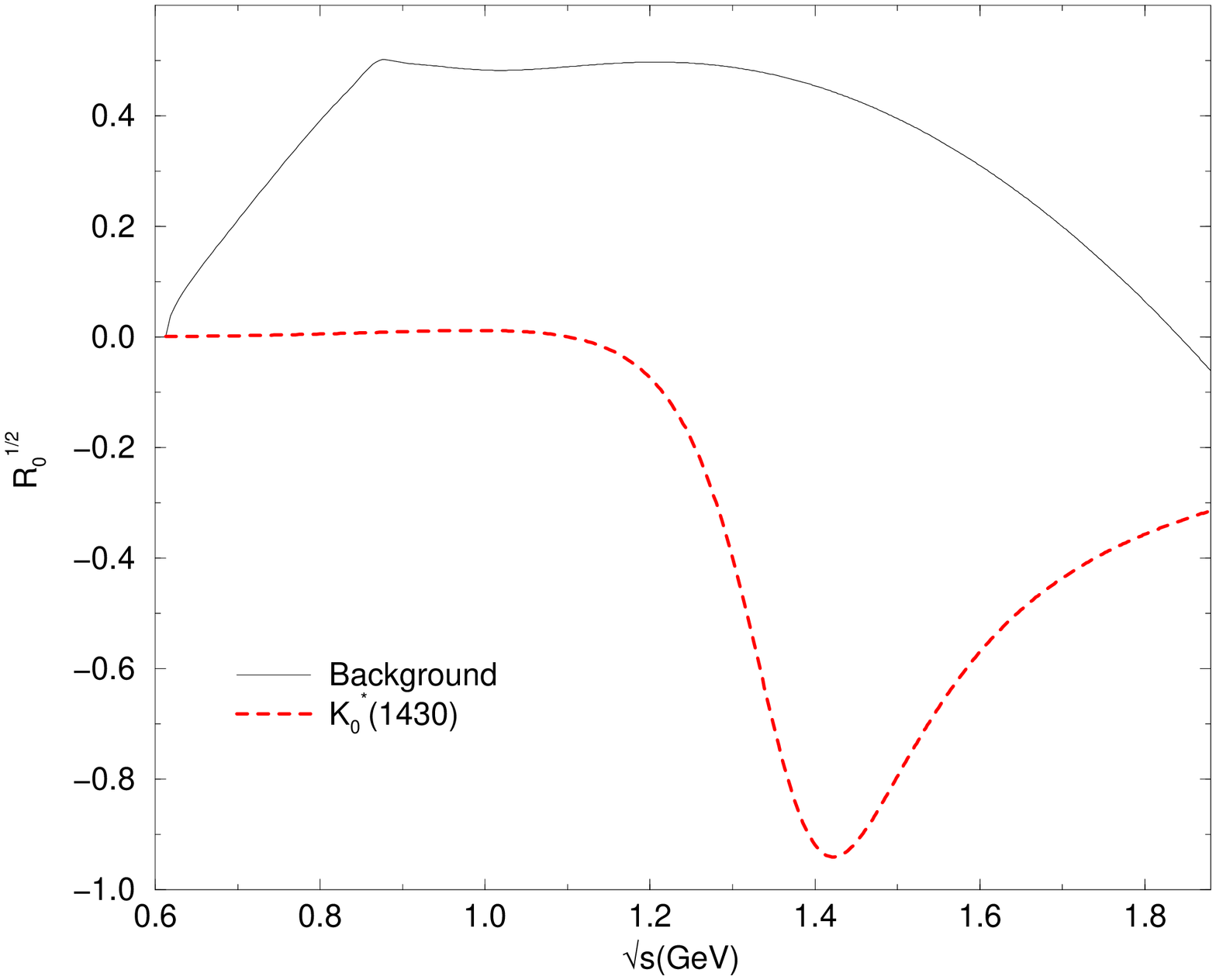, height=5in, angle=270}
\caption
{
Separate contributions of the background and $K^*_{0}(1430)$ to  
$R^{1/2}_0$ (for choice $\gamma_{\sigma K\bar K}=\gamma_{\sigma\pi\pi}$).
}
\label{fig-BGand1430}  
\end{figure}

We now continue with a more quantitative approach in order to extract
the physical parameters of the $\kappa$ meson and the $K_0^*(1430)$.
We fit the theoretical amplitude, which consists \footnote{We also
included the $f_0(1300)$ contribution, which is however very small.}
of the real part of the partial wave projection of (\ref{K(1430)-1/2})
added to $\widetilde R_0^{1/2}(s)$, defined above, to the experimental
data displayed in Fig.~{\ref{fig-expdata}}.  The parameters to be fit
are the three quantities $m_\kappa$, $\gamma_{\kappa K\pi}$ and
$G'_\kappa$ for the $\kappa$ (see Eq. (\ref{kappa-1/2})) and the
corresponding quantities for the $K_0^*(1430)$, namely $m_*$,
$\gamma_*$ and $G'_*$ (see Eq. (\ref{K(1430)-1/2})).  As discussed at
the end of section II, it seems reasonable to obtain the three
$K_0^*(1430)$ parameters self-consistently from our model rather than
taking them from \cite{Aston}.  The scalar meson coupling constants
listed in (\ref{couplings}) were used while, in light of its
uncertainty, the calculation was performed for a range of values of
$\gamma_{\sigma K\bar K}$.  The fitting procedure made use of the
MINUIT package and the fitted parameters, together with their $\chi
^2$ values, are shown in Table \ref{gsKK_Res_Fit}.  It is interesting
to notice that the fitted parameters vary smoothly with
$\gamma_{\sigma K\bar K}$.  The actual comparison between experiment
and the fitted amplitude, using the parameters from the first column
in Table \ref{gsKK_Res_Fit}, is shown in Fig.~{\ref{fig-fit}}.  The
individual contributions due to the background and to the
$K_0^*(1430)$ are shown in Fig.~{\ref{fig-BGand1430}}, indicating that
the background does not violate the unitarity bound at $s=m_*^2$.  The
exact value of the phase found in this fit is ${\rm sin}2\delta = 0.937$.
This agrees with the qualitative discussion regarding the background
phase at the beginning of this section.
\begin{table}   
\begin{center}
\begin{tabular}{|c|c|c|c|}
Fitted Parameter&  $\gamma_{\sigma K{\bar K}}=\gamma_{\sigma\pi\pi}$ 
                &  $\gamma_{\sigma K{\bar K}}=0$  
                &  $\gamma_{\sigma K{\bar K}}=-\gamma_{\sigma\pi\pi}$ 
 \\  \hline
$m_\kappa$ &   897 $\pm$ 2.1 MeV  
           &   951 $\pm$ 0.7 MeV 
           &   998 $\pm$ 1.1 MeV
\\ \hline
$G'_\kappa$  & 322 $\pm$ 6.0 MeV
                 & 277 $\pm$ 10.6 MeV
                 & 195 $\pm$ 5.3 MeV 
\\ \hline
$\gamma_{\kappa K\pi}$  & 5.0 $\pm$ 0.07 GeV $^{-1}$ 
            & 4.32 $\pm$ 0.16 GeV $^{-1}$
            & 4.04 $\pm$ 0.08 GeV $^{-1}$
\\ \hline
$m_*$  & 1385 $\pm$ 3.3 MeV
                 & 1365 $\pm$ 2.5 MeV
                 & 1349 $\pm$ 2.1 MeV 
\\ \hline
$G'_*$ &  266 $\pm$ 9.5 MeV
     &  201 $\pm$ 9.8 MeV
     &  148 $\pm$ 5.6 MeV 
\\ \hline
$\gamma_*$  & 4.3  $\pm$ 2.1 GeV $^{-1}$
                 & 3.7  $\pm$ .1 GeV $^{-1}$
                 & 3.1  $\pm$ 0.05 GeV $^{-1}$
\\ \hline
$\chi^2$  &    4.0
          &    9.0 
          &   25.7
\end{tabular}
\end{center}   
\caption
{ 
Comparison of different fits in the $J=0$ $\displaystyle{I=\frac{1}{2}}$ channel, corresponding to
different choices of $\gamma_{\sigma K{\bar K}}$.
}
\label{gsKK_Res_Fit}
\end{table}

The partial decay width of $K^*_0(1430)$ can be calculated using
(\ref{K(1430) partial width}).  We find that $\Gamma \left(
K^*_0(1430) \rightarrow {\pi K} \right)= 238$ MeV and as a result
(identifying $G'_*$ as the total width) an estimate of the branching ratio
of $K^*_{0}(1430)$ to decay to $\pi K$ can be made
\begin{equation}
{\rm B}\left[ K^*_{0}(1430)\rightarrow \pi K \right] = 
{
 {\Gamma_{K^*_{0}(1430)}}
  \over
  {G'_*}
}
=0.895.
\end{equation}
This quantity is comparable to the 0.93 obtained in \cite{Aston}.
Similarly, the (first column of Table \ref{gsKK_Res_Fit}) mass and width we obtain -
$1385$ MeV and $266$ MeV - are in reasonable agreement with their
\cite{Aston} respective values - $1429$ MeV and $287$ MeV.

\section{$J=0$, $\displaystyle{I=\frac{3}{2}}$ Channel}
\begin{figure}
\centering
\epsfig{file=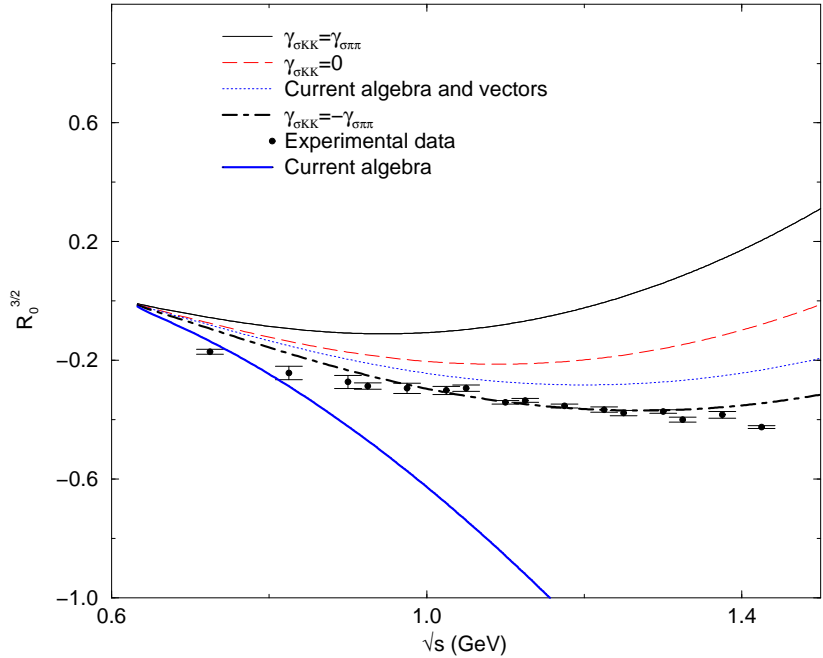, height=5in, angle=270}
\caption
{
Comparison of various predictions for $R^{3/2}_0$ with experiment.
}
\label{fig-R3_2}
\end{figure}

It is interesting to compare with experiment the projection into the
$J=0$, $\displaystyle{I=\frac{3}{2}}$ channel of the same invariant
amplitude used for the last section.  The structures of the invariant
$\displaystyle{I=\frac {3}{2}}$ amplitudes may actually be read off
from Eqs. (\ref{CA-3/2}) - (\ref{sigma-3/2}) of Appendix C.  Since
there are no $\displaystyle{I=\frac{3}{2}}$ resonances in our model,
there are no s-channel poles, and hence this calculation depends
little on the details of the regularizations.  As in the
$\displaystyle{I=\frac{1}{2}}$ case, cancellations of individual
contributions to the partial wave amplitude act to preserve the
unitarity bound.  The experimental points for the real part
$R_0^{3/2}$ were translated from Fig.~12 of \cite{Estabrooks} and are
displayed in our Fig.~\ref{fig-R3_2}.

Fig.~\ref{fig-R3_2} also shows various predictions from our model.
Firstly, we see that the current algebra prediction alone quite soon
departs from the data points and begins to violate the unitarity bound
at around $900$ MeV.  Inclusion of the $\rho$, $K^*$ and contact
contributions associated with the vector mesons can be seen to pull
the curve up considerably so as to solve the unitarity problem and to
give a much better fit to the data.  This is very analogous to the
situation in the $J=0$, $I=2$ partial wave for $\pi\pi$ scattering
(see Fig.~4 of \cite{Sannino-Schechter} and Fig.~ 2.10 of
\cite{Francesco's thesis}).  At this stage, the curve does not depend
on any unknown parameters.

It turns out that the only additional important contribution to this
channel comes from $\sigma$ meson exchange.  This will depend on the
choice of the coupling constant $\gamma_{\sigma K\bar K}$ which was
the important unknown parameter in the previous section.
Fig.~\ref{fig-R3_2} shows the results for the three choices of
$\gamma_{\sigma K\bar K}$ given in Table \ref{gsKK_Res_Fit}.  The best
choice for the $\displaystyle{I=\frac{3}{2}}$ amplitude is the case
$\gamma_{\sigma K\bar K} = -\gamma_{\sigma \pi\pi}$ which
unfortunately yields the fit with the highest $\chi ^2$ for the
$\displaystyle{I=\frac {1}{2}}$ analysis.  The small difference between
the curve for $\gamma_{\sigma K\bar K} = 0 $ and the curve for the
current algebra plus vector contribution measures the small impact of
the other scalars.  Actually the general trend of the data is
reproduced for all values of $\gamma_{\sigma K\bar K}$ shown.

Since there are no large direct channel resonance contributions, the
$\displaystyle{I=\frac{3}{2}}$ amplitude may be especially sensitive
to exchanged resonances in the range above $1$ GeV which we are
currently neglecting.  This is in contrast to the $\displaystyle{I=
\frac {1}{2}}$ amplitude which contains fitting parameters that can
absorb the effects of higher resonance exchanges.  This was the case
for the $\pi\pi$ scattering calculation also.

As we lower $\gamma_{\sigma K{\bar K}}$, we find fits with
larger values of $\chi^2$ that correspond to a $\kappa$ that is heavier,
narrower, and has larger coupling constant, and to a $K^*_{0}(1430)$ that is 
lighter, narrower, and has larger coupling constant.

\section{Discussion}
We have found that a large $N_c$ motivated approximate treatment of
$\pi K$ scattering can give a crossing symmetric and unitary amplitude
as a fit to the existing experimental data.  A novel feature of this
approach, which is analogous to that employed for $\pi\pi$ scattering
in \cite{Sannino-Schechter,Harada-Sannino-Schechter}, is to start with
the invariant perturbative amplitude which is manifestly crossing
symmetric.  This results in individual contributions dramatically
violating the partial wave unitarity bounds.  We rely on cancellations
among these competing contributions to rescue unitarity.  In our
framework this suggests the existence of a light strange scalar
resonance $\kappa$ which has parameters mass $m_{\kappa} = 897$ MeV
and width $G_{\kappa} = 322$ MeV.  These give a pole position
\begin{equation}
\left( s_{\kappa} \right) ^{1/2} = (0.911 - 0.158i) {\rm GeV}.
\label{kappa pole}
\end{equation}
We do not quote any error here since the main uncertainty in this
analysis is clearly due to the theoretical model.  It is noteworthy
that these results are similar to those of \cite{Ishida_kappa} in
which a different model was employed.  In addition, the fit for the
$K_0^*(1430)$ properties also obtained is similiar to that of the
experimental analysis of \cite{Aston}.  Our work was simplified by
directly making use of the analogous approximation seen to be
reasonable in \cite{Harada-Sannino-Schechter} for the $\pi\pi$
scattering case.  Thus, as suggested by working to leading order in
$\displaystyle{\frac {1}{N_c}}$, we compared the real part of the
partial wave amplitude with experiment.  Since elastic unitarity seems
\cite{Aston} to be a reasonable approximation until about the
$K^*_0(1430)$ region for the $\displaystyle{J=0, I=\frac {1}{2}}$
partial wave amplitude, we can recover its imaginary part as
\begin{equation}
I_0^{1/2} \approx \frac {1}{2} \left[ 1 \pm \sqrt { \left(
{\eta}_0^{1/2} \right)^2 - 4 \left( R_0^{1/2} \right)^2} \right],
\end{equation}
with ${\eta}_0^{1/2} \approx 1$ and an appropriate choice of sign.  Of
course the phase shift is recovered as
$\displaystyle{{\rm tan}({\delta}_0^{1/2}) = \frac {I_0^{1/2}}{R_0^{1/2}}}$.

As in the $\pi\pi$ treatment we neglected, for an initial analysis,
the contributions of most resonances above $1$ GeV.  Specifically, we
did not include diagrams with the radially excited vectors $\rho
(1450)$ and $K^*(1420)$ or with the tensors $f_2(1270)$ and
$K_2^*(1430)$.  In a ``second generation'' treatment of this problem
it would be desirable to fully investigate these aspects.  It would be
amusing to see if the complicated $1-2$ GeV region is high enough so
that the ``microscopic'' approach we are following merges with a kind
of string picture \cite{Veneziano}.  

If one accepts the existence of the $\kappa {(900)}$ and $\sigma
{(560)}$, in addition to the $f_0(980)$ and $a_0(980)$, then there is
a full set of candidates for a possibly unconventional (i.e. not of
pure $q \bar q$ type) low mass scalar nonet.  The nature of such a
nonet is of great interest - see \cite{EFSS} for a recent discussion.
A useful clue may arise from knowledge of the pattern of
$0^{+}0^{-}0^{-}$ coupling constants defined in Eqs. (\ref{sigma-int})
- (\ref{K-int}).  The numerical values obtained in our approach are
given in Eq. (\ref{couplings}) and Table \ref{gsKK_Res_Fit}.

 \acknowledgments
The work of D.B., A.H.F. and J.S. has been 
supported in part by the US DOE under contract
DE-FG-02-85ER 40231.
The work of F.S. has been partially 
supported by the US DOE  under contract 
DE-FG-02-92ER-40704.

\appendix

\section{Scattering Kinematics}
The partial wave scattering matrix for 
a channel like $\pi K \rightarrow \pi K$ can be  
written as
\begin{equation}
S=1 + 2 i T \ ,
\label{scattering}
\end{equation}
where for simplicity the isospin and the 
angular momentum variables have not 
been indicated. The standard parameterization of the 
single-channel scattering amplitude is 
\begin{equation}
S=\eta e^{2i\delta_{\pi K}} \ ,
\label{phase}
\end{equation}
where $\delta_{\pi K}$ is the phase shift and 
$0<\eta \leq 1$ is the elasticity parameter. 
Evidently, 
\begin{equation}
T^I_l (s)=\frac{\eta^I_l (s) e^{2 i \delta^I_{\pi K; l}  (s)}  - 1  
}{2 i} \ ,
\end{equation}
where $l$ and $I$ label the angular momentum 
and isospin, respectively.  
The real and imaginary parts 
\begin{equation} \displaystyle
R^{I}_{l}=\frac{\eta ^I_l \sin \left( 2 \delta^I_{\pi K; l}\right)}{2}\  
,
\quad\quad I^I_{l}=
\frac{1 - \eta ^I_l \cos \left( 2 \delta^I_{\pi K; l}\right)}{2} \ ,
\label{real-imaginary}
\end{equation}
must satisfy the very important unitarity bounds
\begin{equation}
\left|{R^I_{l}} \right| \leq \frac{1}{2} \ , 
\quad \quad 0\leq I^I_{l} \leq 1 \ .
\label{unitarity}
\end{equation}
Now we relate the previous partial wave amplitudes to the 
$\displaystyle{I=\frac{1}{2}}$ and 
$\displaystyle{I=\frac{3}{2}}$ invariant amplitudes for 
the scattering process $\pi (p_1) + K(p_2) \rightarrow \pi (p_3) + K  
(p_4)$.  
This is simply achieved by first defining the 
$\displaystyle{I=\frac{3}{2}}$ amplitude via 
\begin{equation}
A^{3/2} \left( s,t,u \right) = A\left( \pi ^+ (p_1) K^+(p_2)
\rightarrow \pi ^+ (p_3) K^+(p_4) \right) \ ,
\label{3/2}
\end{equation}
where s, t and u are the Mandelstam variables.  By crossing symmetry
we have $\displaystyle{A\left( \pi ^+ K^- \rightarrow \pi ^+ K^-
\right) = A^{3/2} \left(u,t,s\right)}$ which leads to
\begin{equation}
A^{1/2}\left(s,t,u\right)=\frac{3}{2} A^{3/2}\left(u,t,s\right) - 
\frac{1}{2} A^{3/2} (s,t,u) \ .
\label{1/2}
\end{equation}
We then define the partial wave isospin 
amplitudes according to the formula
\begin{equation}
T^I_{l} (s) = \frac{\rho(s)}{2} \int^1_{-1} d\!\cos \theta 
\, P_l (\cos \theta) \, A^{I} (s,t,u) \ ,
\label{projected}
\end{equation} 
where $\theta$ is the scattering angle and 
\begin{equation}
\rho(s) = \frac {q(s)}{8\pi \sqrt s} \equiv \frac{1}{16 \pi \, s} 
\sqrt{\left[ s - \left(m_{\pi} + m_{K}\right)^2 \right]   
\left[ s - \left(m_{\pi} - m_{K}\right)^2 \right] } \ .
\label{kinematical}
\end{equation}

\section{Notation and Lagrangian}
Spontaneous chiral symmetry breaking plays a fundamental role at low
energies and is often economically as well as successfully described
by non-linear realizations.  Associated with the standard chiral
symmetry breaking pattern $SU(3)_{\rm L}\times SU(3)_{\rm R}
\rightarrow SU(3)_{\rm V}$ we have an octet of pseudoscalar
Nambu-Goldstone bosons $\phi$.  The latter are encoded in a 3 $\times$
3 matrix $U$ as follows,
\begin{equation}
U = \xi^2 \ , \qquad \xi = e^{ i\frac{\phi}{F_\pi}} \ ,
\end{equation}
where $F_\pi$ is the pion decay constant.
$U$ transforms under a chiral transformation as  
\begin{equation}
U \rightarrow U_{\rm L} U U_{\rm R}^{\dag} \ ,
\label{trans: U}
\end{equation}
with $U_{\rm L,R} \in \mbox{U(3)}_{\rm L,R}$.
While $U$ transforms linearly 
under these transformations 
(see Eq.~(\ref{trans: U})),
$\xi$ transforms non-linearly, i.e.
\begin{equation}
\xi \rightarrow 
U_{\rm L} \, \xi \, K^{\dag}(\phi,U_{\rm L},U_{\rm R}) = 
K(\phi,U_{\rm L},U_{\rm R}) \, \xi \, U_{\rm R}^{\dag} \ .
\end{equation}
The vector meson nonet $\rho_\mu$ may be formally introduced as a 
{\it gauge field} \cite{Kaymakcalan-Schechter}.
It transforms under chiral rotations as 
\begin{equation}
\rho_\mu \rightarrow K \rho_\mu K^{\dag} + 
\frac{i}{\widetilde{g}} K \partial_\mu K^{\dag} \ ,
\end{equation}
where $\widetilde{g}$ is the {\it gauge coupling constant}.
(For an alternative approach see, for a review, 
Ref.~\cite{Bando-Kugo-Yamawaki:PRep}.)
It is convenient to define the following  objects 
\begin{eqnarray}
p_\mu &=& \frac{i}{2}
\left( 
  \xi \partial_\mu \xi^{\dag} - \xi^{\dag} \partial_\mu \xi
\right) \ ,
\nonumber\\
v_\mu &=& \frac{i}{2}
\left( 
  \xi \partial_\mu \xi^{\dag} + \xi^{\dag} \partial_\mu \xi
\right) \ ,
\end{eqnarray}
which obey the transformation rules
\begin{eqnarray}
p_\mu &\rightarrow& K p_\mu K^{\dag} \ , \nonumber\\
v_\mu &\rightarrow& K v_\mu K^{\dag} + i K \partial_\mu K^{\dag} \ .
\end{eqnarray}
Using the above quantities
we can construct  the non-anomalous part of the chiral 
Lagrangian describing pseudoscalar and vector mesons:
\begin{equation}
{\cal L} =
-\frac{1}{2} m_v^2 \mbox{Tr} 
\left[ \left(\rho_\mu - \frac {v_\mu}{\widetilde{g}} \right)^2 \right]
- \frac{F_\pi^2}{2} \mbox{Tr} 
\left[ p_\mu p_\mu \right]
-\frac{1}{4} \mbox{Tr} 
\left[ F_{\mu\nu}(\rho) F_{\mu\nu}(\rho) \right],
\label{Lag: sym}
\end{equation}
where
$F_{\mu\nu} = \partial_\mu \rho_\nu - 
\partial_\nu \rho_\mu - i \widetilde{g} 
[ \rho_\mu , \rho_\nu ]$
is the vector meson {\it gauge field strength}.
Chiral symmetry is explicitly broken in QCD by 
the presence of an explicit quark mass term 
$- \widehat{m} \overline{q} {\cal M} q$,
where $\widehat{m} \equiv (m_u+m_d)/2$,
and ${\cal M}$ is the dimensionless matrix:
\begin{equation}
{\cal M} = \left(
\begin{array}{ccc}
1+y & & \\ & 1-y & \\ & & x
\end{array} \right) \ .
\end{equation}
Here $x$ and $y$ are the quark mass ratios:
\begin{equation}
x = \frac{m_s}{\widehat{m}} \ , \qquad
y = \frac{1}{2}
\left(\frac{m_d-m_u}{\widehat{m}}\right) \ .
\end{equation}
These quark masses induce a mass term for 
the pseudoscalar mesons which at the 
effective lagrangian level is represented 
by the following term 
\begin{equation}
{\cal L}_{\phi{\rm-mass}} = \delta' \mbox{Tr} \left[ {\cal M} U^{\dag}
+ {\cal M}^{\dag} U \right] \ ,
\label{pi mass 1}
\end{equation}
where $\delta'$ is a real constant.  A more general set of terms
describing explicit chiral symmetry breaking in this framework is
available in
Refs.~\cite{Schechter-Subbaraman-Weigel,Harada-Schechter}.  Scalar
resonances, in the non linear realization framework, interact with
pseudoscalars with at least two derivatives. If we were to identify
the scalars with a matter field nonet, i.e.. which transforms under
chiral transformations as $S\rightarrow K S K^{\dagger}$ a possible
invariant interaction term is $\displaystyle{{\rm Tr} \left[ S p_{\mu}
p_{\mu}\right]}$.  Since the quark content of the scalars is not yet
firmly established and other possible terms may exist we adopt here a
more phenomcnological approach by not relating the scalar couplings
using SU(3) symmetry.  For the present paper, the relevant
interaction terms are

\begin{eqnarray}
{\cal L}_{\sigma}&=& -\frac{\gamma_{\sigma \pi \pi}}{\sqrt{2}} 
\sigma \partial_{\mu} \mbox{\boldmath ${\pi}$}\cdot
\partial_{\mu}{\mbox{\boldmath ${\pi}$}} -
\frac{\gamma_{\sigma K \bar K}}{\sqrt{2}} \sigma \left( 
\partial_{\mu}K^{+} \partial_{\mu}K^{-} + .... \right) \ ,  
\label{sigma-int} \\
{\cal L}_{f_0}&=& -\frac{\gamma_{f_0\pi \pi}}{\sqrt{2}} 
f_0 \partial_{\mu} \mbox{\boldmath ${\pi}$}  \cdot \partial_{\mu} 
\mbox{\boldmath ${\pi}$} -
\frac{\gamma_{f_0 K\bar K}}{\sqrt{2}} f_0 \left( 
\partial_{\mu}K^{+} \partial_{\mu}K^{-} + .... \right) \ ,  
\label{f0-int} \\
{\cal L}_{\kappa} &=& - {\gamma_{\kappa K\pi}} \left( 
\kappa^0\partial_{\mu} K^{-} \partial_{\mu} \pi^+ + {....} \right) \ .
\label{K-int}
\end{eqnarray}
Different models will relate the coupling constants in different ways.
For example in the $SU(3)$ limit, and if the scalars belong to the
usual matter field nonet with no OZI violating interactions, we have
$\displaystyle{\gamma_{\sigma \pi \pi} = \gamma_{\sigma K\bar K}=
\frac{\gamma_{f_0 K\bar K}}{\sqrt{2}}=\gamma_{\kappa K \pi}}$ while
$\gamma_{f_0 \pi \pi} =0$.

\section{Unregularized Amplitudes}
The current-algebra contribution to the 
$A^{3/2}\left(s,t,u\right)$ amplitude, obtained from (\ref{Lag:
sym}) and from (\ref{pi mass 1}) is:
\begin{equation}
A^{3/2}_{CA}\left(s,t,u\right)=\frac{t+u-s}{2F^2_{\pi}} \ .
\label{CA-3/2}
\end{equation}
The vector meson contribution contains the following terms 
\begin{eqnarray}
A^{3/2}_{vect}\left(s,t,u\right)&=&
\frac{g_{\rho \pi \pi}^2}{4} 
\left[\frac{u-s}{m^2_{\rho} - t} 
-  \frac{m^2_{K^*} \left(s - t \right)  - 
{\left(m^2_K - m^2_{\pi}\right)^2}}
{\left(m^2_{K^*} - u\right) {m^2_{K^*}} } \right] \nonumber \\
&+&\frac{g^2_{\rho \pi \pi}}{4 m^2_{\rho}} \left(2s - u - t \right) \  
,
\label{vect-3/2}
\end{eqnarray}
where $\displaystyle{g_{\rho \pi \pi}=\frac{m^2_{\rho}} {\widetilde{g}
F^2_{\pi}}}$ is the coupling of the vector to two pions, which is
related to the width by $\displaystyle{\Gamma \left( \rho \rightarrow
2\pi \right) = \frac {g^2_{\rho\pi\pi} p^3_\pi}{12 \pi m^2_{\rho}}}$.
The first and second terms correspond respectively to $\rho^0$ and
$K^*$ exchanges, while the third term represents the contact
interaction $v_\mu v_\mu$ in (\ref{Lag: sym}).  The contribution of a
strange scalar, denoted $\kappa$, is
\begin{equation}
A^{3/2}_{\kappa}\left(s,t,u\right)=\frac{\gamma^2_{\kappa K\pi}}{4} 
\frac{\left( u - m^2_{\pi} - m^2_{K}\right)^2}{m^2_{\kappa} - u} \ . 
\label{kappa-3/2}
\end{equation}
Finally the $\sigma$ exchange contribution is
\begin{equation}
A^{3/2}_{\sigma}\left(s,t,u\right)=
\frac{\gamma_{\sigma \pi \pi} 
\gamma_{\sigma K \bar K}}{4} 
\frac{\left(2 m^2_{\pi} - t\right) 
\left(2 m^2_{K} - t\right)}{m^2_{\sigma} - t} \ .
\label{sigma-3/2}
 \end{equation} Note that (\ref{kappa-3/2}) can also be used to
describe the contribution of the scalar resonance $K^*_{0}
\left(1430\right)$, if we reidentify the coupling constant and the
mass in the denominator.  Similarly (\ref{sigma-3/2}) can be used for
the $f_0$ exchange if we replace each subscript $\sigma$ by a
subscript $f_0$.  

The $A^{1/2}$ amplitudes are obtained from these
using (\ref{1/2}).

\end{document}